\documentclass[%
twocolumn,
 longbibliography,
 amsmath,amssymb,
 aps,
 pre,
]{revtex4}

\usepackage{mathtools}
\usepackage{lipsum}
\usepackage{graphicx}
\usepackage{dcolumn}
\usepackage{bm}
\usepackage{comment}
\usepackage{afterpage}
\usepackage{xcolor}
\definecolor{sapphire}{HTML}{0067A5}
\usepackage[colorlinks=true,citecolor=sapphire,linkcolor=sapphire,urlcolor=sapphire]{hyperref}

\begin{document}
\title{The shape of cleaved tethered membranes}
\author{A. D. Chen$^1$, M. C. Gandikota$^{1,2}$ and A. Cacciuto$^1$}
\email{ac2822@columbia.edu}
\affiliation{$^1$ Department of Chemistry, Columbia University\\ 3000 Broadway, New York, NY 10027\\
$^2$ International Centre for Theoretical Sciences, Tata Institute of Fundamental Research, Bengaluru 560089, India}

\begin{abstract}
A remarkable property of flexible self-avoiding elastic surfaces (membranes) is that they remain flat at all temperatures, even in the absence of a bending rigidity or in the presence of active fluctuations. 
Here, we report numerical results of these surfaces wherein we alter their topology by systematically cleaving internal bonds. While it is known that a random removal of membrane bonds does not disrupt the overall extended shape of the membrane, we find that cleaving an elastic surface with longitudinal parallel cuts leads to its systematic collapse into a number of complex morphologies that can be controlled by altering the number and length of the inserted cuts. For the simpler case of membranes with bending rigidity but in the absence of self-avoidance, we find that the radius of gyration of the surface as a function of number of cuts is represented by a universal master curve when the variables are appropriately rescaled.
\end{abstract}

\maketitle
\section{Introduction}

Membranes are $D$-dimensional structures embedded in a higher $d$-dimensional space. \textit{Tethered} membranes (also referred to as crystalline/polymerized membranes)~\cite{kantor1986,bowick2001statistical,gompper1997network,NelsonBook,wiese2000,bowick2001} are characterized by the fixed connectivity of their constituent nodes, in contrast to fluid membranes, which do not have such explicit connectivity and undergo neighbor exchange~\cite{NelsonBook}. The energetics of tethered membranes comprise of a stretching energy for the elastic (tethering) bonds, a bending energy between neighboring regions accounting for the thickness of the material, and an excluded volume term that models surface self-avoidance. 
Physical tethered membranes are ubiquitous both in synthetic and biological materials such as graphite-oxide~\cite{wen1992crumpled}, graphene sheets~\cite{Stankovich2006Jul}, polymer films~\cite{Huang2007Aug}, cytoskeleton of red blood cells~\cite{Schmidt1993Feb,Lux2016Jan} and viral capsids~\cite{Lidmar2003Nov}. 

When compared to their 1-dimensional (1D) polymer counterparts,  2D tethered ideal membranes display rather unique shape properties~\cite{wiese2000,bowick2001}. In \textit{ideal} membranes (i.e., membranes without self-avoidance), the interplay between thermal and bending forces, characterized by the bending rigidity, $\kappa$, determine the equilibrium state of the surface. These are known to exhibit a second-order phase transition between a flat (high $\kappa$) and a crumpled phase (low $\kappa$), with a critical bending rigidity of $\kappa_{\text{c}}\simeq 0.33\,k_{\text{B}}T$ ~\cite{kantor1986}. 
In the crumpled phase, the radius of gyration of the system depends logarithmically on the linear size of the membrane $L$~\cite{NelsonBook}. 
The stability of the flat phase is rationalized as the outcome of the non-linear coupling between the in-plane and out-of-plane modes of deformation of the membrane, which renormalizes its bending rigidity leading to a stiffening of the surface with system size~\cite{peliti1987,aronovitz1988,chaikin1995}. 

\begin{figure*}[t]
    \centering
    \includegraphics[width=0.9\linewidth]{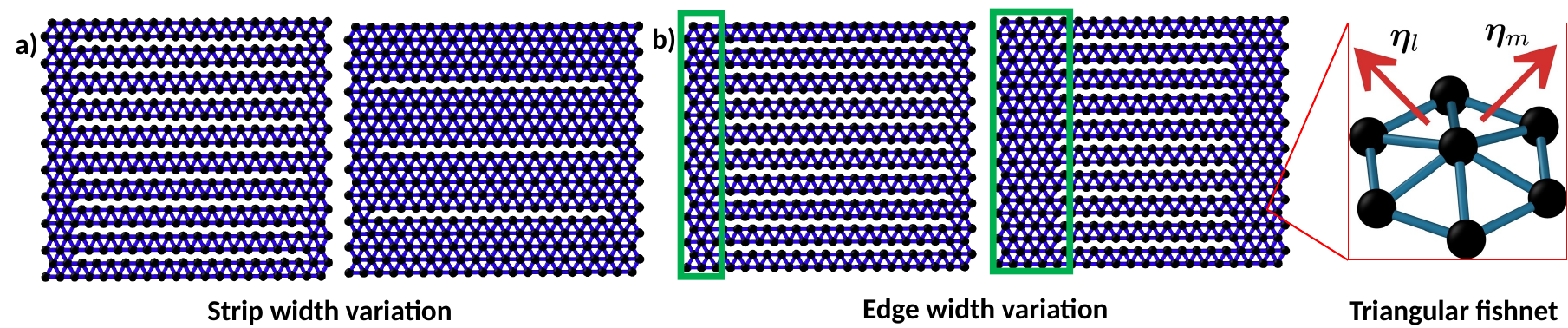}
    \caption{A sketch of the cleaved membrane model consisting of nodes and bonds (a) Cleaved membranes with different numbers of parallel cuts (horizontal). (b) Highlight of the edge-widths, $d_e$, which determines the length of the parallel cuts. Inset: An element in the triangular fishnet network defining the membrane. The variable $\bm{\eta}$ indicates the normal to each triangle from which the bending forces are computed.}
    \label{fig:scheme}
\end{figure*}
Physical membranes however, obey self-avoidance. For a \textit{self-avoiding} tethered membrane of linear dimension $L$ in a good solvent, a simple Flory-type argument predicts a high temperature Flory phase for which the radius of gyration scales as $R_{\text{g}}\sim L^{\nu}$ with the size exponent $\nu=(D+2)/(d+2)$~\cite{wiese2000}. For two-dimensional  membranes ($D=2$) in a three-dimensional embedding space ($d=3$), $\nu= 4/5$, and the fractal (Hausdorff) dimension $D_{\text{f}}= 2.5$~\cite{domb2001,paczuski1988}.  
Numerical simulations of self-avoiding tethered membranes, have shown, however,  that tethered membranes remain strictly extended, irrespective of the temperature of the thermal bath~\cite{plischke1988,abraham1989,bowick1996flat}. This result has been supported by variety of numerical models, including a triangulated network of hard spheres imposing several different degrees of self-avoidance~\cite{boal1989,abraham1990} and a plaquette model, where the hard spheres are removed, and self-avoidance is imposed by explicitly preventing triangle-triangle intersections among the discrete parts of the membrane surface~\cite{bowick2001}. 

One way to destabilize the flat phase consists of placing the surface in a poor solvent. Indeed, using this method leads to a multi-step folding transition through a variety of folded states until the surface eventually collapses into compact shape ($\nu=2/3$, $D_{\text{f}}=3$)~\cite{abraham1991folding}. 
Alternatively, the flat phase can be destabilized via an external energy input, where the crumpling process is essentially of a non-equilibrium nature. An example of such processes at the macroscopic scale includes the crumpled phase obtained simply by crushing sheets of aluminium foil, for which a scaling exponent of $\nu \approx 4/5$ was observed~\cite{kantor1987}. At the microscale, a crumpled phase is observed when aerosolized water droplets containing graphene-oxide nanosheets are passed through a furnace for rapid dehydration~\cite{ma2012}. Force-compression curves of self-avoiding sheets have also been numerically well studied~\cite{vliegenthart2006forced}. More recently, it has been shown that while active fluctuations play a role analogous to temperature in open tethered membranes~\cite{gandikota2023}, tethered shells crumple even for small activities~\cite{Gandikota2024-in}.

Overall, it is fair to say that, in the absence of explicit external non-equilibrium forces or effective attractive interactions between its elements, a floppy self-avoiding tethered membrane will remain stubbornly flat, limiting its possible engineering applications. In this paper, we show how a number of new surface morphologies can be achieved  by performing parallel cuts across its surface, in what is essentially
a process of systematic surface bond dilution.
 Earlier works~\cite{Grest1990-fo,plischke1991monte} have shown that random removal of the monomer and bond from the surface of a self-avoiding tethered membrane fails to disrupt its overall flat shape even when a very high proportion of monomers are randomly removed, to the point where the membrane begins to fall apart. Therefore, it is important to perform judicious and systematic cuts on the surface. The number of cuts and their length across the surface are the key architectural parameters that we use to alter the shape of the surface. 

Other work on systematic bond removal obtained by creating periodic membrane perforations~\cite{yllanes2017thermal} has also been reported. However, this work dealt exclusively with ideal membranes and showed that the presence of the perforations leads to a lower critical temperature for the surface crumpling transition. 

\section{Model and methods}

We model the surface as a triangular fishnet of beads and springs embedded in three dimensions
(Fig. \ref{fig:scheme}a). Each bead is represented as a spherical  particle of diameter $\sigma$. The surface has an overall square shape with a linear side length (number of particles per side) $L$. In this study, we considered membranes with $L$ in the range $24-100$ corresponding to a total number of nodes $N$ equal to $576-10000$, respectively. The membranes are cleaved into parallel strips of width $w$ by removing bonds between the monomers of adjacent strips (Fig. \ref{fig:scheme}a). The number of cuts in the surface, $n_c$, and the length of the membrane, $L$, are related to the width of the membrane strips, $w$, via $w=L/(n_c-1)$. 
Various values of $w$ are used, all integer factors of the side length $L$ of that given membrane (Fig. \ref{fig:scheme}b).
We begin and terminate cuts at various distances from the edges of the membrane. This distance we refer to as the edge-width, $d_e$. This is related to the membrane side length via $d_e=(L-\ell_{cut})/2$, where $\ell_{cut}$ is the length of each cut (Fig. \ref{fig:scheme}b). Through these systematic cuts, we seek to encourage membrane deformations and drive the surface towards non-flat conformations.

We study three model systems: ideal cleaved membranes without bending rigidity, ideal cleaved membranes with small bending rigidities, and self-avoiding cleaved membranes without bending rigidity. The dynamics of the system is determined by the Langevin equation, 
\begin{equation}
    m\frac{d\bm{v}_i}{dt} = \bm{f}_i - \gamma\,\bm{v}_i + \sqrt{2D\gamma^2}\, \bm{\xi}_i(t)
\end{equation}
where $m$ is the mass of any given node particle, $i$ represents the index of a given particle. The velocity of each particle is $\bm{v}_i$, and $\gamma$ is the translational friction. The conservative forces on each given particle $\bm{f}_i$ are found by differentiating the interaction potential $V$ (detailed below) with respect to the particle position $\bm{r}_i$: $\bm{f}_i = -\frac{\partial V}{\partial \bm{r}_i}$. $D = k_{\text{B}}T\gamma^{-1}$ is the translational diffusion constant, where $k_{\text{B}}$ is the Boltzmann constant, and $T$ is the temperature. The interaction potential of the system is,
\begin{equation}
\begin{split}
        V &= K\sum_{\langle ij \rangle}^{\text{neighbors}}(r_{ij}-r_{\rm eq})^2 + \kappa\sum_{\langle lm \rangle}^{\text{neighbors}}(1-\bm{\eta}_l\cdot\bm{\eta}_m) \\
        &+4\epsilon\sum_{ij}\left[\left(\frac{\sigma}{r_{ij}}\right)^{12}-\left(\frac{\sigma}{r_{ij}}\right)^{6}+\frac{1}{4}\right],
\end{split}
\end{equation}
where $r_{ij}$ is the distance between any given pair of nodes. The first term represents the harmonic bonding energy term between neighboring particles. The equilibrium bond length is set to be $r_{\rm eq} =1.6\,\sigma$ with a spring constant $K$, which was set to $K=160k_{\text{B}}T/\sigma^2$. The second term represents the bending energy. Here, $\kappa$ is the bending rigidity, and $\bm{\eta}_l$ and $\bm{\eta}_m$ represent the vectors normal to any pair of adjacent triangles of the surface. 
The last term is a repulsive Weeks-Chandler-Anderson (WCA) potential\cite{weeks_role_1971} that is cut off at $2^{1/6}\sigma$ to enforce volume exclusion for each particle, with $\epsilon = k_{\text{B}}T$. These simulations have been carried out using the numerical packages LAMMPS~\cite{Plimpton1995-rq} and HOOMD blue~\cite{Glaser2015-mc}. In these simulations, the units of length, time, and energy are set to be $\sigma$, $\tau=\sigma^2D^{-1}$, and $k_{\text{B}}T$, respectively. The timesteps used in these simulations were of length $\Delta t = 0.015 \tau$. Given these dimensionless units, we set $\gamma=1$.

To characterize the shapes that cleaved membranes of different topologies may exhibit, we use the shape tensor~\cite{rudnick_aspherity_1986},
\begin{equation}
    S_{\alpha\beta} = \frac{1}{2N^2}\sum^N_{i=1}\sum^N_{j=1}\;(r_{i\alpha}-r_{j\alpha})(r_{i\beta}-r_{j\beta})
    \label{eq:shape}
\end{equation}
where $\alpha$ and $\beta$ are the Cartesian coordinates of the 3D system ${x,y,z}$, and $i$ and $j$ represent particle indices. This shape tensor accounts for the products of different Cartesian displacements between all particles within the system. Diagonalizing this tensor results in three eigenvalues associated with three principle directions. These eigenvalues will be denoted as $\lambda_{\alpha}$, where $\alpha =$ 1, 2, or 3. The relative sizes of the eigenvalues are sorted such that $\lambda_{\text{1}} \geq \lambda_{\text{2}} \geq \lambda_{\text{3}}$, where each eigenvalue can be conceptualized as the average displacement of the shape away from its center of mass along the associated principle axis of the shape, with the first axis being the largest and the third axis being the shortest. For a 1D rod, we would expect $\lambda_{\text{1}}$ to proportional to the length of the rod  while $\lambda_{\text{2}} = \lambda_{\text{3}} = 0$. For an infinitely thin, flat square, we would expect $\lambda_{\text{1}} = \lambda_{\text{2}}$ and these to also be proportional to the lateral size of the surface, while $\lambda_3 = 0$. The sum of the three  eigenvalues is the overall squared radius of gyration of the shape, i.e., $\lambda_{\text{1}} + \lambda_{\text{2}} + \lambda_{\text{3}} = R_{\text{g}}^2$.

\section{Ideal membranes}
\subsection{Zero bending rigidity}\label{ideal_zero}
\begin{figure}[t]
    \centering
    \includegraphics[scale=0.26]{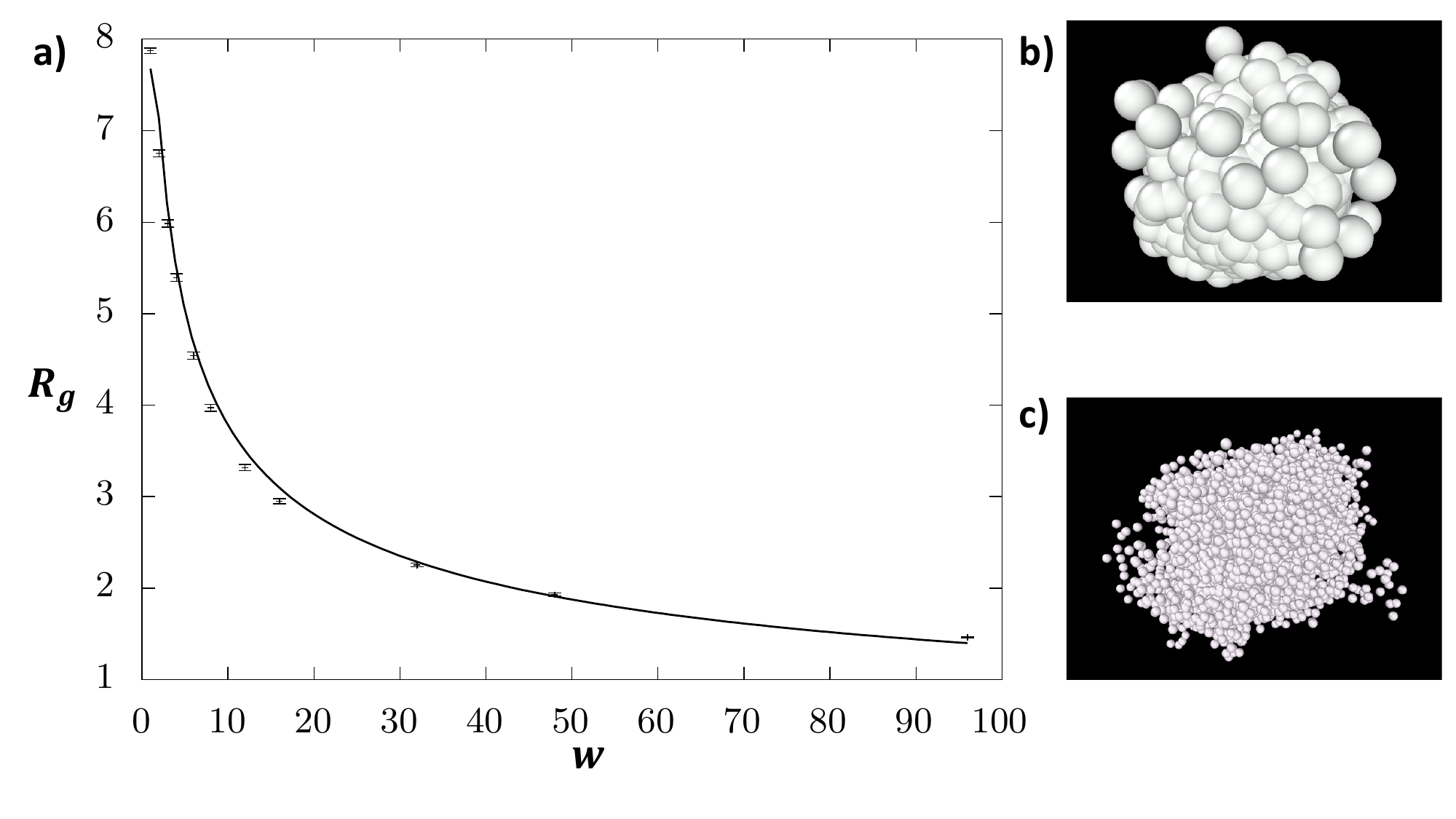}
    \caption{Shape of ideal cleaved membranes with zero bending rigidity. a) The radius of gyration, $R_{\text{g}}$, decreases  as a function of strip-width $w$. Here, the system size is $L=96$ and the edge-width is $d_e=1$. The line is a fit to the data according to the functional form $R_{\rm g}=\alpha(L/w)^{1/2}(1+\beta\sqrt{\text{log}(w)})$, with $\alpha \simeq 0.78(2)$ and $\beta \simeq 0.37(3)$. Snapshots of the membrane from simulations for $w=1$ and $w=L$ are shown respectively in (b) and (c).
    }
    \label{fig:idnbrrg}
\end{figure}

We begin our investigation considering the simplest case: an ideal  membrane without bending rigidity (i.e., $\epsilon=0$ and $\kappa =0$). For this system, we considered a fixed edge-width, $d_e=\sigma$. Under these conditions, the intact (un-cleaved) membrane is in a highly crumpled phase. 
We cleave the membrane by performing periodic longitudinal cuts across the surface, which determine the strip-widths $w$. The resultant morphological changes in the membrane are characterized using radius of gyration, $R_g$, as the order parameter, which is seen to monotonically decrease with increasing $w$ (see Fig.~\ref{fig:idnbrrg}). In other words, the more cuts the membrane has (i.e., smaller $w$), the larger is its size. 
This behaviour can be understood by considering the two limits of the cleaved membrane, $w=L$ which is an intact, uncleaved membrane and $w=1$ ($L-1$ cuts along the surface) which is essentially a bundle of polymers connected by a square frame on the edges. The compact volume of the intact membrane can be expected to follow a size-scaling $R_g\sim \sqrt{\log(L)}$~\cite{NelsonBook} whereas in the other limit, the bundle of polymers can be expected to follow $R_g\sim L^{1/2}$~\cite{Gennes1979Nov}. Thus, in the thermodynamic limit, the bundle of polymers (which has a power-law size-scaling) will be much larger than the intact membrane with a logarithmic scaling.
In both cases, the membrane is fairly isotropic.
We find that the crossover between these two limiting cases can be precisely described by a simple function of the form 
$R_{\text{g}} = \alpha(L/w)^{1/2}(1+\beta\sqrt{\text{log}(w)})$, where $\alpha\simeq 0.78(2)$ and $\beta \simeq 0.37(3)$ are model-dependent constants for $L = 96$ (Fig.~\ref{fig:idnbrrg}). This expression is an interpolation between the two limits which reduces to $R_{\text{g}} \sim \sqrt{\log(L)}$ when $L = w$ and to $R_{\text{g}} \sim L^{1/2}$ when $w=1$.

\subsection{Finite bending rigidity}
\begin{figure}[t]
    \centering
    \includegraphics[width = 0.48\textwidth]{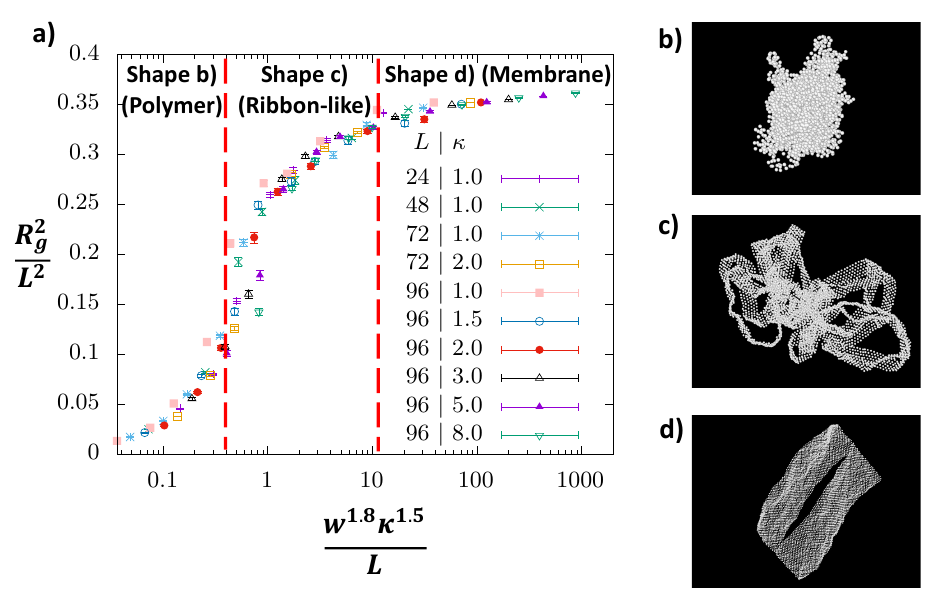}
    \caption{Shape of ideal cleaved membranes with finite bending rigidity. (a) The collapse of the rescaled radius of gyration as a function of $w^{1.8}\kappa^{1.5}/L$ for different system-sizes $L$, bending constants $\kappa$ and strip-widths $w$. In all cases, the edge-width $d_e$ is set to be 2. 
    Dashed lines represent rough phase boundaries. Labels within the plot correspond to representative snapshots of morphologies for $L = 72$ shown in (b), (c), and (d), where $w=$ 2, 6, and 36, respectively.}
    \label{fig:idbrmaster}
\end{figure} 

\begin{figure*}[t]
    \centering
    \includegraphics[width=0.9\linewidth]{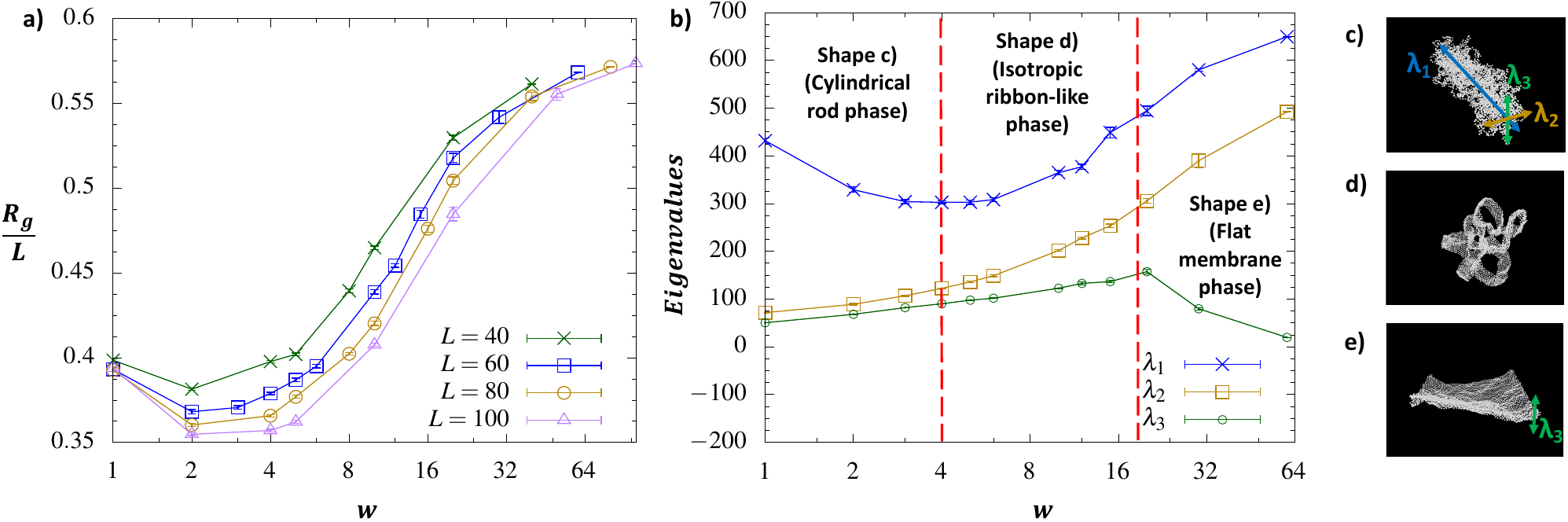}
    \caption{Shape of self-avoiding cleaved membranes with zero bending rigidity. (a) The radius of gyration displays a minimum as a function of strip-width $w$ for all values of $L$ studied, as seen in this linear-log plot. (b) The eigenvalues of the shape tensor for an $L = 60$ membrane as a function of $w$. Dashed lines represent rough phase boundaries. Labels within the plot correspond to representative snapshots of morphologies shown in (c), (d), and (e). For all data above, edge-width $d_e$ is set to one.}
    \label{fig:sanbrrg}
\end{figure*} 
Next, we introduce a finite bending rigidity $\kappa$ to these ideal membranes.
Specifically, we consider values of $\kappa$ above the critical point, $\kappa_{\text{c}} = 0.33\,k_{\text{B}}T$, so that in the limit of $w\rightarrow L$, where the membrane is intact~\cite{NelsonBook}, we recover, as the equilibrium reference shape, a flat membrane. 
We study this system for a variety of membrane lateral lengths $L$ and bending rigidities $\kappa$ as a function of $w$ for $d_e=2\sigma$. 
We find an interesting variety of diverse shapes that the membranes systematically adopt with increasing number of cleavages (see Fig. \ref{fig:idbrmaster}b-d). 
In contrast with Sec.~\ref{ideal_zero}, the $R_{\rm g}$ is seen to \textit{increase} with $w$. This is a consequence of the now finite bending constants $\kappa>\kappa_\text{c}$ where the full, intact membrane is in the flat phase, while for $w=2$, the system still exists as a bundle of polymers, albeit with a slightly larger persistence length. 
In fact, for $w = 2$, we retain the size-scaling of the radius of gyration to be $R_{\rm g}\sim L^{0.51(1)}$, as confirmed by a power law fit (Fig. S1), and for $w=L$, the scaling is $R_{\rm g}\sim L$. Upon increasing $w$, the cleaved membrane expands until it eventually flattens out. The expansion of the membrane can be qualitatively understood by recognizing that each strip is a ribbon whose rigidity increases with increasing $w$.

Remarkably, within the range of sizes and rigidities considered in this work, all the data obtained for different membrane side-lengths $L$ and bending rigidities $\kappa$ collapse onto a master curve where $(R_{\rm g}/L)^2$ is seen to be a function of $w^{1.8}\kappa^{1.5}/L$ (see Fig. \ref{fig:idbrmaster}a). 
The overall plateau of the master curve for large $w$ is a consequence of the linear scaling of  $R_{\text{g}}$ with $L$. The initial linear growth for $w\gtrsim 2$ comes from the scaling of the membrane $R_{\text{g}}$ in the polymer bundle limit, $L^{1/2}$, when plotted against $1/L$ in the horizontal axis.
Since we were able to map the behavior in the large and small $w$ limits to ideal membranes and polymers, respectively, it is tempting to attempt to map the behavior in the intermediate $w$ regime to analogous ideal systems. From the snapshot of the membrane in this region depicted by Fig. \ref{fig:idbrmaster}c, we can see that the cleaved membrane resembles a cluster of ideal ribbons tethered together. In this case, a ribbon can be considered as a two-dimensional object with an aspect ratio interpolating between that of a 1D polymer and a 2D tethered membrane. In ring-like elastic frames, which are essentially four ribbons connected end-to-end, the rescaled order parameter $R_{\text{g}}^{2}/L^{2}$ is a function of $l_{\text{p}}/L = w^{1.8}\kappa^{0.2}/L$~\cite{yllanes2017thermal,yllanes2019folding}. The renormalized persistence length $l_{\text{p}}$ of a ribbon accounts for thermal fluctuations~\cite{kovsmrlj2016response,yllanes2017thermal}.
Attempting to collapse our data using this scaling variable does not produce satisfactory results (Fig. S2). While the $w$ dependence is similar to that which we empirically found for our cleaved membrane system, the $\kappa$ dependence is rather different. This suggests that cleaved membranes are unlike free ideal ribbons for intermediate strip-widths $w$, indicating the importance of tethering at the edges in determining the morphology of cleaved membranes.

\section{Self-avoiding Membranes}
Lastly, we study the effect of the parallel cuts on the shape of self-avoiding tethered membranes in the absence of bending rigidity, and draw a diagram capturing the different morphologies in terms of strip width $w$ and edge width $d_e$. 
As mentioned earlier, self-avoidance alone is sufficient to drive a tethered membrane into a flat phase~\cite{NelsonBook}. For a $d_e=1$, we observe a distinct non-monotonic behavior of the radius of gyration as a function of strip-width (Fig.~\ref{fig:sanbrrg}a). The non-monotonicity becomes progressively more pronounced with increasing system-size. 
We can see that when we rescale $R_{\text{g}}$ by $L$, the points for larger values of  $w$ (i.e., $w = 40$ for $L = 80$ or $w = 50$ for $L = 100$) fall at roughly the same value on the y-axis (apart from systematic finite size effects), implying that $R_{\text{g}}/L \sim {\text{const.}}$, consistent with a flat shape, for which $R_{\text{g}}\sim L$.

More interesting is the $w\rightarrow 1$ limit, for which all rescaled curves converge to roughly the same value, which suggests that a linear dependence of $R_{\text{g}}$ on the membrane lateral length $L$ also holds in the polymer limit ($w=1$). Crucially, this scaling is not indicative of a flat phase, but rather of an elongated, rod-like morphology (see Fig. \ref{fig:sanbrrg}c). 
To further characterize the shape of the membrane, we compute the shape tensor~(eq. \ref{eq:shape}), the eigenvalues of which at various $w$ of an $L = 60$ membrane system are shown in Fig. \ref{fig:sanbrrg}b.

We can see in the plot of Fig.~\ref{fig:sanbrrg}b that $\lambda_1 \gg \lambda_2 \approx \lambda_3$ in the small $w$ regime, indicating a rod-like shape, with the longest axis parallel to the eigenvector whose eigenvalue is $\lambda_1$ (Fig. \ref{fig:sanbrrg}c). 
In the small $w$ regime, $\lambda_1$ decreases with increasing $w$ while $\lambda_2$ and $\lambda_3$ increase systematically their value. 
In the intermediate $w$ range, $\lambda_1$ reverses its trend with $w$, and $\lambda_2$ and $\lambda_3$ begin to diverge from each other as they continue to increase.
As shown in the snapshot in Fig.~\ref{fig:sanbrrg}c, this represents a more isotropic phase, comprised of many ribbons tethered together at the edges to form a beach-ball-like structure. The degree of isotropy within this phase can also be confirmed by monitoring the asphericity parameter defined as~\cite{rudnick_aspherity_1986},
\begin{equation}
A=\frac{3}{2}\frac{\lambda_1^2+\lambda_2^2+\lambda_3^2}{(\lambda_1+\lambda_2+\lambda_3)^2}-\frac{1}{2},
\end{equation}
which presents a minimum in this region (see Fig.~\ref{fig:asph}).
Upon a further increase of $w$, $\lambda_1$ and $\lambda_2$ continue to grow, while the value of $\lambda_3$ begins to drop off drastically. As can be seen from the shape in Fig.~\ref{fig:sanbrrg}b,e this corresponds to the cleaved membrane adopting the shape of a flat surface with $\lambda_1 \approx \lambda_2 \gg \lambda_3$, where $\lambda_3$ corresponds to the out-of-plane fluctuations of this membrane. 

Clearly, the initial decay of $R_g$ with $w$ (for small $w$) can then be traced back to the corresponding decay of the first eigenvalue in that regime. The decay of $\lambda_1$ suggests that as we increase $w$, the cylindrical shape becomes less rigid along its main axis ($\lambda_1$). 
To understand this behavior, one should recognize that in this regime, the cleaved surface behaves essentially as a bottle-brush polymer, where the edge of the surface acts as the polymer backbone while the strips act as the side chains whose excluded volume interactions stabilize the overall cylindrical shape. In the absence of an intrinsic bending rigidity of the backbone, the persistence length of a bottle-brush increases with the grafting density of the side-chains~\cite{Mohammadi2021Nov,Paturej2016Nov,Sunday2023Sep,Feuz2005Oct,Chatterjee2016Nov}. In our setup, an increase in $w$ is equivalent to a decrease in this grafting density, resulting in a decrease of bottle-brush persistence length. This is reflected in the decrease of $\lambda_1$.  
As $w$ further increases, the strips become wider. The theory of ribbons suggests that they acquire an effective persistence length which grows with $w$ and become more rigid~\cite{Giomi2010,kovsmrlj2016response,yllanes2017thermal}.
This increased stiffness with $w$ pushes the edges of the surface further apart, thus increasing $\lambda_2$ and $\lambda_3$ as the surface approaches the isotropic expanded ribbon shape. 
This overall isotropic shape develops as the strips cross over from polymers to ribbons. 
Further increase of $w$ causes the ribbons to become even stiffer, initially isotropically expanding the isotropic structure, but eventually flattening the cleaved membrane out to an extended shape.

We also find that the initial decay of the $R_g$ becomes less pronounced and eventually vanishes as one increases the edge-width, $d_e$, of the membranes, effectively making the strips shorter.
Fig.~\ref{fig:sanbredge} show the results of this analysis for edge-widths $2\,\sigma$ and $4\,\sigma$. 
Beyond the overall shape of the curve, the more important effect of increasing $d_e$ is that of suppressing the intermediate isotropic phase. 
Indeed, for sufficiently large values of $d_e$, the edges of the membrane become stiffer, and already for small values of $w$, we can no longer easily conceptualize these cleaved membranes within the context of bottle-brush polymers. 

\begin{figure}[t]
    \centering
    \includegraphics[width = 0.48\textwidth]{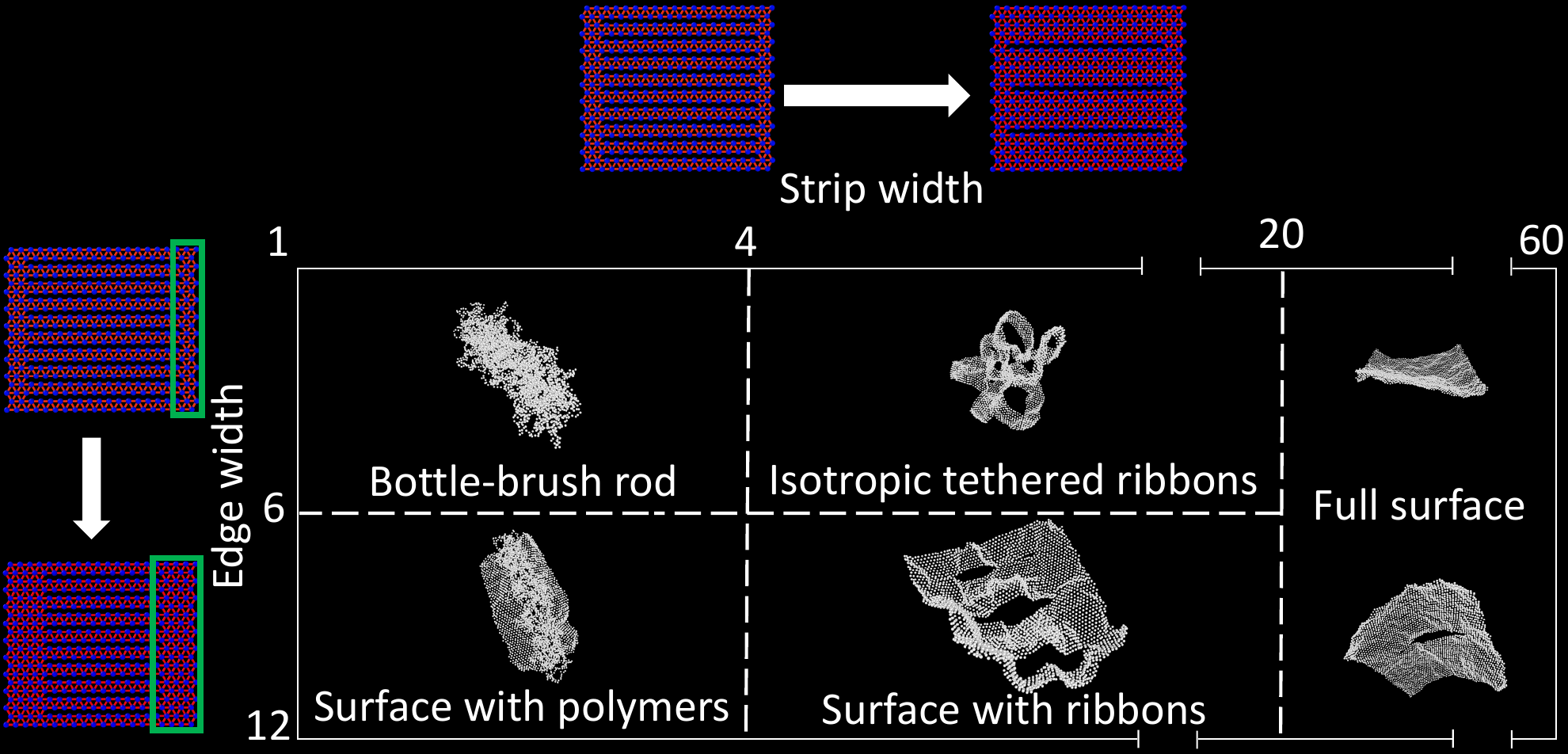}
    \caption{Qualitative sketch of structural diagram of the self-avoiding cleaved membranes. The phases of these membranes can be tuned using strip-width $w$ (x-axis) and edge-width (y-axis) as control parameters. Dashed lines represent rough shape boundaries.}
    \label{fig:phase}
\end{figure}

Figure~\ref{fig:phase} recapitulates the diverse shape landscape of the cleaved membranes as a function of $w$ and $d_e$. For this study we considered $d_e\in [\sigma,\,12\sigma]$, with the membranes depicted in Fig.~\ref{fig:phase} being $L = 60$. At the bottom of the figure, on the l.h.s., we find a conformation that can be described as two co-planar flat surfaces (the edges) opposite to each other connected by a tangle of thin, flexible strips. In this scenario ($w\rightarrow 1$, $d_e\gg 1$), the effective rigidity of the surface is primarily due to the effective stiffness of the edges, rather than to the self-avoidance of the strips, the latter mostly contributing to maintain the co-planarity of the edges. 
Increasing $w$ while keeping $d_e\gg 1$ does not significantly affect the size of $\lambda_1$, but has the effect of pushing the membrane edges apart from each other, effectively expanding the surface along the edge's planes, and bypassing the formation of the isotropic ribbon shape. We find this boundary to occur at around $d_e \approx 6\sigma$ for an $L = 60$ membrane.

\section{Conclusions}
Previously, numerical work on tethered surfaces has shown that self-avoiding elastic membranes tend to stay in a flat phase even in the absence of bending rigidity. In this paper, we suggest a simple method to alter this structure by performing finite, parallel cuts across the surface. Cleaving a micro- or nano-scale surface is experimentally possible today because of a number of technological advances that have occurred over the previous decade. For instance, nano-ribbons can now be created by cutting graphene nanosheets into strips down to sub 10 nm precision using methods such as artificial design patterning combined with hydrogen plasma etching~\cite{Shi2011-gu}. Non-trivial zig-zag patterns can also be obtained, highlighting the precision and utility of such techniques to generate desired patterns (see reference ~\cite{Lou2024-cu} for a comprehensive review of such techniques).
In this paper, we find that judicious control of number of cuts and their lengths on the surface of a membrane allows for significant control over the shape it can achieve, including access to an isotropic non-compact phase. The shapes acquired by the cleaved membranes are determined by a delicate balance between the effective stiffness of the edges versus the stiffness and number of lateral strips formed by the parallel cuts. We computed a structural diagram enumerating the different shapes acquired by a self-avoiding membrane under different conditions, and discussed in detail how this behavior changes when considering ideal membranes (in the absence of excluded volume effects) with an explicit bending rigidity. Here, we found that upon increasing the number of cuts, the surface moves from a flat shape to an expanded isotropic structure, until it eventually collapses into a shape---the size of which is well described by the radius of gyration of a single polymer. Conversely, in the absence of an explicit bending rigidity, adding cuts to the ideal surface leads to an increase of the overall size of the membrane, albeit no flat shape can ever be formed.

Interestingly, for ideal membranes, we find that all data can be collapsed, when appropriately rescaled, into a  single master curve. While the dependence on $w$ is compatible with that of renormalization group calculations for individual strips~\cite{kovsmrlj2016response}, we find a rather different dependence on the bending rigidity.
Each individual strip should follow the power law predicted in ~\cite{kovsmrlj2016response}, so that an increase in $w$ would correspond to an increase in the stiffness of each individual strip, resulting in the expansion of the membrane as a whole. However, it is unknown how the renormalization of the bending rigidity for the entire membrane would be affected by the presence of longitudinal cuts and the external rim connecting the strips. What appears clear is that the cuts soften the overall rigidity of the membrane because of the systematic decrease in the number of bonds on the surface,  thus requiring a stronger dependence on $\kappa$ to achieve the same stiffness of an intact membrane.

It is apparent that the formation of the different shapes observed in this work is governed by the enhanced flexibility of the ribbons relative to the full, intact membranes, upon introducing membrane cuts. It would be thus important to explore in more detail how the rigidity of these self-avoiding free ribbons, even in the absence of an explicit bending rigidity, depend on their width and/or aspect ratio, to understand how to interpolate between a self-avoiding polymer and a membrane (ongoing work in progress). 

We should also emphasize that the different shapes obtained by cleaving the surface in our system are  driven only by thermal fluctuations, while entropy plays no role in most of the Kirigami tessellations works used to design the shape of elastic manifolds (see~\cite{Kirigami} and references therein). Furthermore, in the latter case,  external forces acting on specific points  need to be applied to drive the surfaces into the desired shapes, while they form spontaneously in our system.
Nonetheless, it would be interesting to explore these sophisticated tessellation strategies in the context of soft thermal surfaces.

\section*{Data availability}
Data for this article, including all data points in the figures and the codes used to run and analyze the trajectories are available at \url{https://drive.google.com/drive/folders/1Qq1auZdRwPnX7mjE6l2OE1ZTlauih_Fd}.

\section*{Acknowledgements}
A.C. acknowledges financial support from the National Science Foundation under Grant No. DMR-2003444.

\bibliography{References}
\bibliographystyle{apsrev4-1}

\section{Appendix}
\renewcommand\thefigure{SI.\arabic{figure}} 
\setcounter{figure}{0}

\begin{figure}[h]
    \centering
    \includegraphics[width=0.9\linewidth]{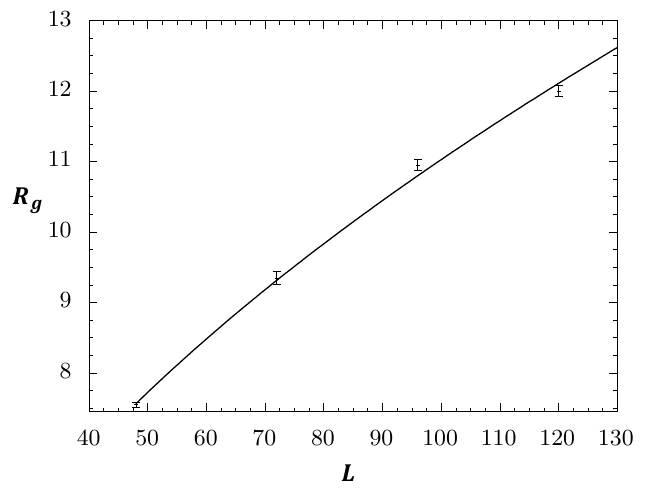}
    \caption{Power law fit of the radius of gyration $R_{\text{g}}$ of an ideal cleaved membrane as a function of the lateral size of the membrane, $L$. The fit has follows the functional form $R_{\text{g}} \sim L^{\nu}$, where $\nu = 0.51(1)$, consistent with ideal polymers. Here, the edge-width $d_e$ and strip-width $w$ were both equal to 2, while $\kappa= k_{\text{B}}T$.}
    \label{fig:powlawfit}
\end{figure} 

\begin{figure*}[t]
    \centering
    \includegraphics[width=0.45\linewidth]{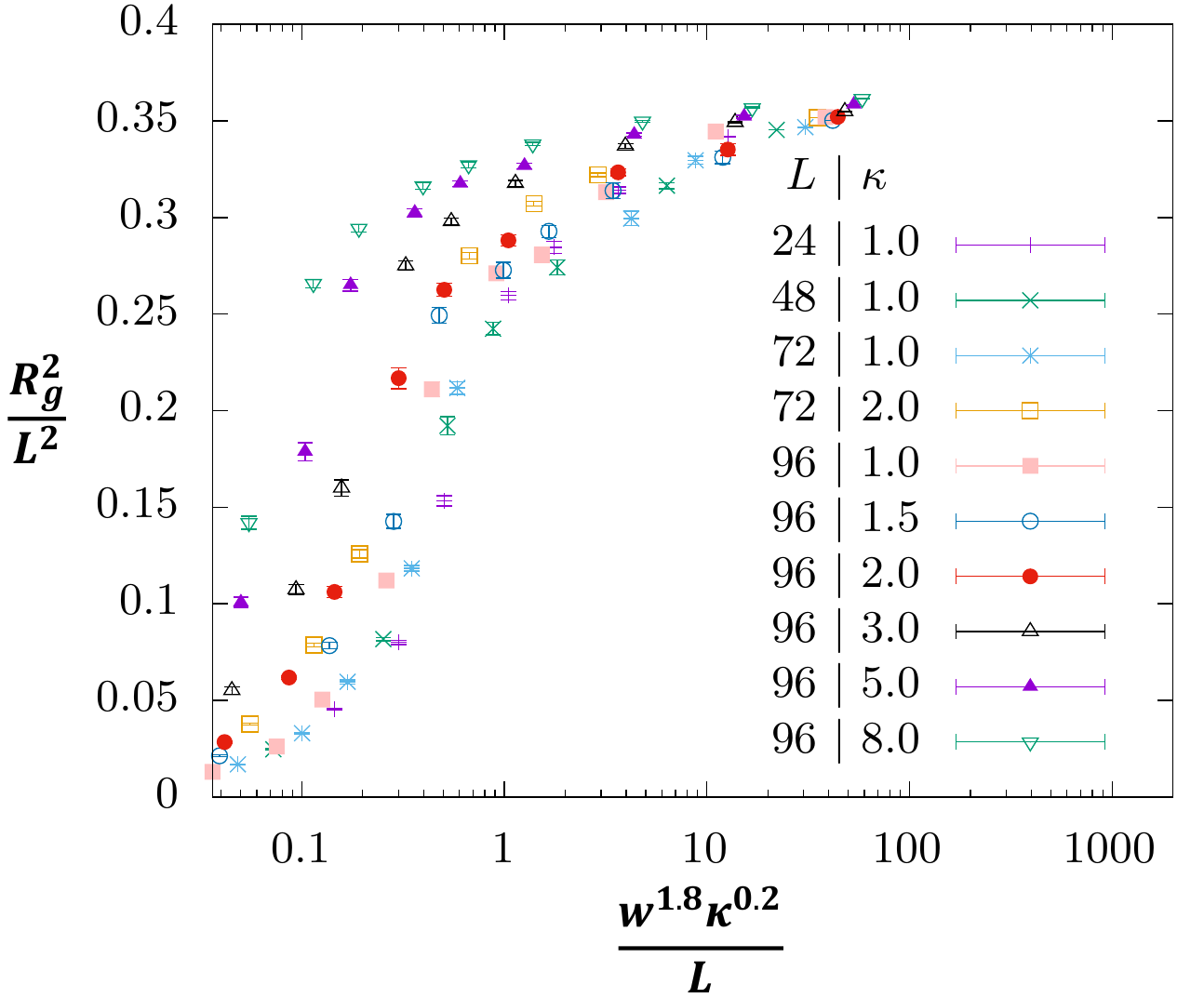}
    \caption{The collapse of the rescaled radius of gyration as a function of $w^{1.8}\kappa^{0.2}/L$ for different system-sizes $L$, bending constants $\kappa$ and strip-widths $w$, as in reference~\cite{yllanes2017thermal}. It should be noted that this scaling parameter with $\kappa^{0.2}$ does not produce as satisfactory of a collapse as compared to when $\kappa^{1.5}$ for our data (see Fig. 3). In all cases, the edge-width $d_e$ is set to be 2.}
    \label{fig:bowick_master}
\end{figure*}

\begin{figure*}[t]
    \centering
    \includegraphics[width=0.45\linewidth]{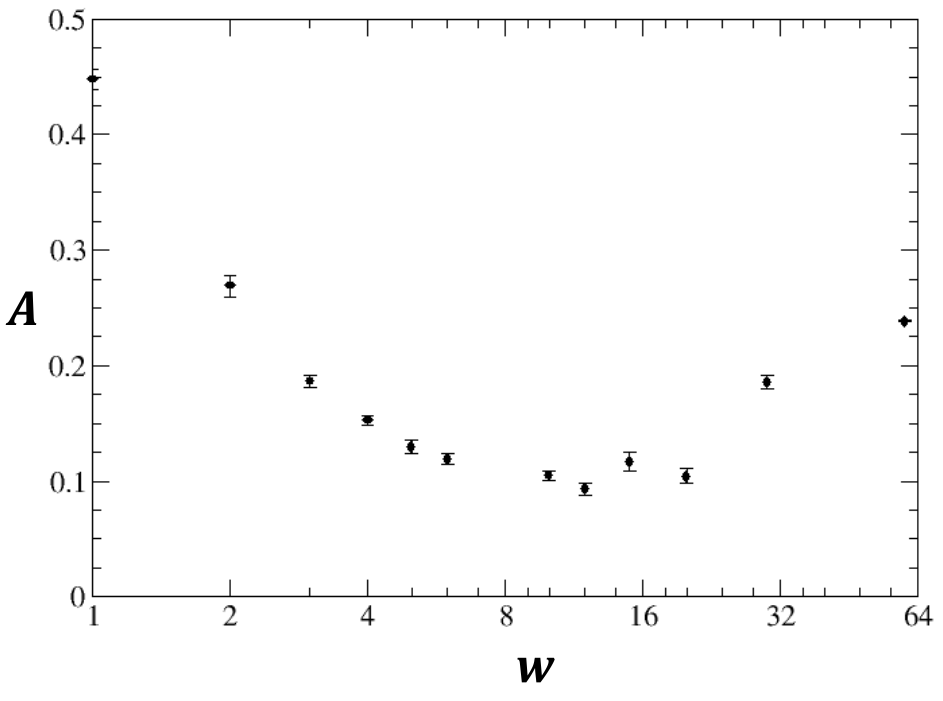}
    \caption{Linear-log plot of self-avoiding cleaved membrane asphericity $A$ as a function of $w$. Here, $L = 60$ and $d_e = 1$.}
    \label{fig:asph}
\end{figure*} 
\begin{figure*}[b]
    \centering
    \includegraphics[width=0.9\linewidth]{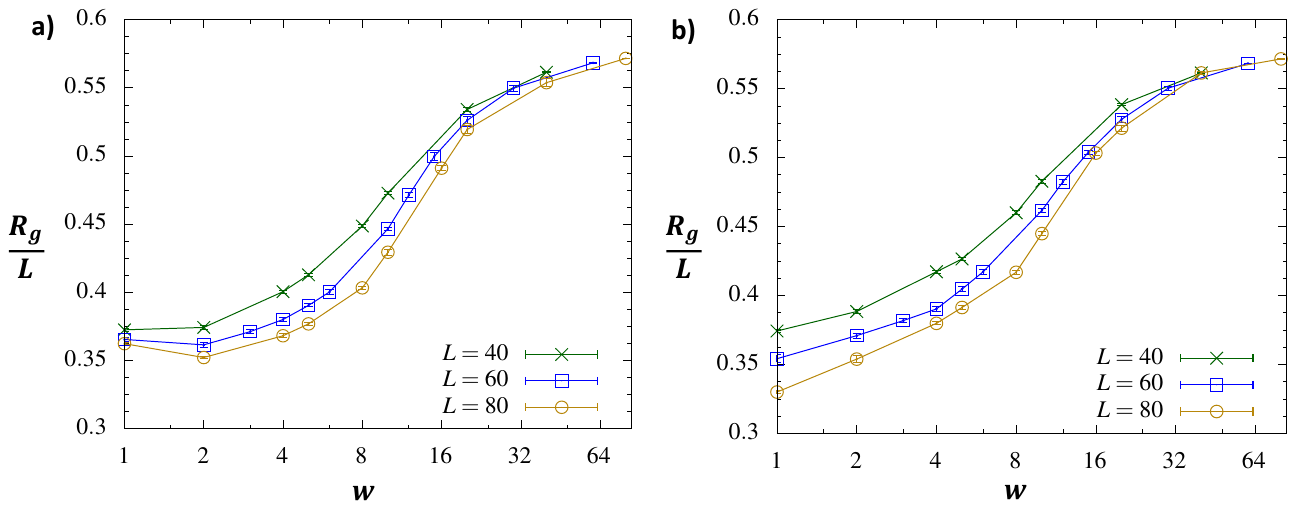}
    \caption{Linear-log plots of $R_{\text{g}}/L$ as a function of the strip width, $w$, for self-avoiding cleaved membranes of different side lengths $L$. The $d_e$ were set to be (a) 2 or (b) 4.}
    \label{fig:sanbredge}
\end{figure*}

\end{document}